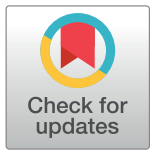

**Data Availability Statement:** All relevant data are within the manuscript and its Supporting Information files.

**Funding:** M.N.: Work partially supported by MIUR, the Italian Ministry of Education, University and Research, under PRIN Project n. 20174LF3T8 AHeAD (Efficient Algorithms for HArnessing Networked Data). A.F.C.: Work partially supported by the University of Perugia, through the program "Fondo Ricerca di Base 2019", project n. RICBA19LTI ("Business Intelligence, Data Analytics e simulazioni a eventi discreti per le Smart Companies nell'era dell'Industria 4.0"). The funders

RESEARCH ARTICLE

# Distinctiveness centrality in social networks

**Andrea Fronzetti Colladon**[1☯]*, **Maurizio Naldi**[2,3☯]

1 Department of Engineering, University of Perugia, Perugia, Italy, 2 Department of Law, Economics, Politics and Modern languages, LUMSA University, Rome, Italy, 3 Department of Civil Engineering and Computer Science, University of Rome Tor Vergata, Rome, Italy

☯ These authors contributed equally to this work.
* andrea.fronzetticolladon@unipg.it

## Abstract

The determination of node centrality is a fundamental topic in social network studies. As an addition to established metrics, which identify central nodes based on their brokerage power, the number and weight of their connections, and the ability to quickly reach all other nodes, we introduce five new measures of Distinctiveness Centrality. These new metrics attribute a higher score to nodes keeping a connection with the network periphery. They penalize links to highly-connected nodes and serve the identification of social actors with more distinctive network ties. We discuss some possible applications and properties of these newly introduced metrics, such as their upper and lower bounds. Distinctiveness centrality provides a viewpoint of centrality alternative to that of established metrics.

## Introduction

The determination of node centrality is a fundamental and popular topic in social network studies [1–3], which never stopped attracting the interest of scholars, e.g. [4–7]. The concept of centrality has been interpreted in many ways, and several metrics have been proposed to study the positional power of social actors [8, 9]. Similarly, different validation approaches were used to assess the role of these metrics in the identification of influential nodes [10]. Three of the most famous centrality metrics—i.e. degree, closeness and betweenness centrality—were described by Freeman [2]. While degree counts how many direct connections a node has, closeness and betweenness are computed considering also indirect connections. Closeness is measured as the reciprocal of the sum of the length of the shortest paths between a node and all other nodes in the graph; it gives an idea of how quickly a social actor can reach its peers. Betweenness centrality counts how many times a node lies in-between the paths that interconnect the other nodes, thus serving as a bridge and acquiring brokerage power.

Other studies introduced the idea that centrality not only depends on the social position of a node but also on that of its neighbours—like in the case of eigenvector centrality [11]. This metric attributes higher scores to nodes connected to other important nodes. "A person with few connections could have a very high eigenvector centrality if those few connections were to very well-connected others" [12]. For example, if a lowly graduate student publishes a paper with her/his supervisor (who has published many papers with others), the student becomes important, simply by virtue of her/his connection to the supervisor. Few connections with extremely important nodes can be enough to make a node important.

On the other hand, scholars like Burt [13] noted that there are cases where social actors exert a stronger influence if their peers are not strongly connected among each other. He posed the question whether having a dense ego-network is beneficial to social capital and showed that individuals might hold positional advantages or disadvantages based on the network they are embedded in (i.e. on the connections among their peers). In particular, missing links among the actors in a node's neighbourhood (structural holes) are often seen as an advantage, as the node can act as a mediator, use a *divide-et-impera* strategy, or combine ideas from different sources and come up with the most innovative one [14]. On the other hand, a high ego-network closure is often seen as a constraint to the brokerage power of the ego, who cannot mediate among its peers. This effect is measured through network constraint, i.e. the extent to which the neighbours of a node are also connected to each other [14]. An alternative metric, based on the same logic, is the effective size one [13, 15], which quantifies the non-redundant part of a person's relationships, with a person's ego-network having redundancy if her/his contacts are connected to each other as well.

Several variations of the above-mentioned metrics were proposed [1, 16], as well as different algorithms for their fast computation on large graphs [17]. Indeed, metrics such as weighted betweenness centrality are costly to compute [18]. However, the majority of centrality metrics tend to attribute stronger influence to nodes that are highly connected, or which are connected to other important nodes. Connections to the network periphery, on the other hand, are often regarded as less relevant.

In this paper, we question this last assumption and propose a new set of metrics—which we call *Distinctiveness Centrality* (DC)—that attribute more importance to nodes which have links to loosely connected others. While we still recognize the pivotal importance of traditional centrality metrics, we also believe that there may be contexts in which connections to peripheral nodes should be valued more. For example, it might be the case that nodes with more peripheral connections keep the network together, avoiding fragmentation. These nodes may be the only ones able to reach certain peers and could be used as a seed for the diffusion of practices that promote health in the population. In other applications, for example when analysing word co-occurrence networks [19] to evaluate brand importance [20], brands with connections to distinctive words may be more important, as they show unique traits that distinguish them from competitors. They convey a different brand image. These are just some examples showing the need for new centrality metrics, which can favour non-redundant connections towards loosely connected nodes. Accordingly, we introduce a new set of indicators that capture the value of distinctive connections and add to the information captured by traditional centrality measures. Distinctiveness centrality is also relatively fast to compute, as it does not require the calculation of shortest network paths, which is necessary for other metrics instead (e.g. closeness and betweenness).

The remainder of the paper is organized as follows: in the next two sections, we define a set of five measures of distinctiveness centrality and compare them with well-known ego-network measures, to show that the information they capture is different. We also derive lower and upper bounds that could be used for normalization, to allow the comparison of scores obtained on different networks. Subsequently, we present the five metrics in the case of directed graphs. In the section named Possible Applications, we provide examples and illustrate some possible use cases. In the last section, we discuss our findings and make proposals for future research.

## Definition of metrics

In this section, we present five metrics of distinctiveness centrality, which were all conceived following the same logic: giving more importance to nodes that are strongly connected to loosely connected peers, so that they make the network periphery more reachable. In the computation of network centrality, all our metrics penalize connections to hubs or nodes that are very well connected. The concept of degree centrality is reinterpreted following this logic.

Let's consider a network that we represent through a weighted undirected graph $G$, which is described by the triplet $G = (V, E, W)$. Let $V$ be the set of nodes of cardinality $|V| = n$, $E = (x, y): x, y \in V, x \neq y$ be the set of arcs, and $W$ be the set of weights associated to the arcs, with $m = \min_W w_{ij}$, $M = \max_W w_{ij}$, $\forall i, j$. If the nodes $i$ and $j$ are not connected, we assume $w_{ij} = 0$, otherwise we assume $w_{ij} \geq 1$. If $m = M$, the graph is practically unweighted; in that case we can rescale all the weights and assume $w_{ij} = 1$.

For the generic node $i \in V$, we introduce five distinctiveness centrality metrics. In the following, $g_j$ is the degree of node $j$ and $I_{(f)}$ is the indicator function which equals 1 if $f$ = TRUE (we will often use the indicator function to account for non-existing arcs). An exponent $\alpha \geq 1$ is used in the formulas to allow a stronger penalization of connections with highly connected nodes. In order not to clutter the notation, the exponent $\alpha$ will not be included as an argument of any metric, though it is clear that the value of metric depends on it.

In the following, we do not consider the case of isolates and compute upper and lower bounds for the new metrics. Indeed, established centrality measures, such as degree, closeness, and betweenness, share a common property of being subject to normalization so that they take values in the [0, 1] range. This property is desirable since it allows to make centrality statements of the low-high kind. Also, it allows comparing the centrality of networks of different sizes. We want this property to hold also for our new centrality measures. Here we limit to the case of connected networks, where no node is isolated so that $g_i \geq 1$, $\forall i$.

**Weighted distinctiveness centrality**

It is defined as

$$D_1(i) = \sum_{\substack{j = 1 \\ j \neq i}}^{n} w_{ij} \log_{10} \frac{n - 1}{g_j^\alpha}. \quad (1)$$

This metric is similar to weighted degree centrality [9], as it sums the weight of all arcs connected to a node. However, weights here are penalized based on the number of connections that a node's peers have. For each node, the sum providing the metric's value is made of as many terms as the degree of the node.

If we set $\alpha = 1$, all the terms are non-negative. However, if a neighbouring node is connected to all the other nodes as well (i.e. $g_j = n - 1$), its contribution to the sum is zero; the rationale is that node $i$ adds the minimum possible improvement to the reachability of node $j$ by connecting it since node $j$ is already connected to all other nodes. Instead, if a neighbouring node is connected to node $i$ only, the weight of the arc $w_{ij}$ connecting them is multiplied by the maximum possible factor $\log_{10}(n - 1)$ (the rationale here is that node $j$ would be unreachable if it were not connected by node $i$).

Instead, if $\alpha > 1$, all the neighbouring nodes whose degree is

$$g_j > e^{\frac{\ln(n - 1)}{\alpha}} \quad (2)$$

provide a negative contribution to the sum and therefore lower the overall value of the metric.

In order to derive the bounds of the metric, we start by considering that the maximum is achieved when all the following conditions are satisfied: a) the node $i$ has the maximum connectivity (so that the sum has the maximum possible number of terms); b) all the neighbouring nodes of node $i$ have minimum connectivity ($g_j$ = 1, $\forall j \neq i$, i.e., they are connected to node $i$ only), which in turn guarantees that all contributions are positive; c) the weights of the arcs connected to the node $i$ are maximum. Under these conditions, we have

$$D_1(i) \leq M(n-1)\log_{10}(n-1). \tag{3}$$

It is to be noted that conditions a) and b) take place if the node $i$ is the hub of a star topology. In addition, we note that the upper bound implied by Eq (3) does not depend on $\alpha$ and is then valid also when $\alpha > 1$.

On the other hand, we get the minimum value of the metric when all contributions are negative and take the largest possible values. Namely, the following conditions have to be satisfied: a) the neighbours of node $i$ are connected to all other nodes ($g_j = n - 1$, $\forall j \neq i$), which in turn leads to negative contributions to the sum for any $\alpha > 1$); b) the node $i$ is connected to all other nodes (which leads to the maximum possible number of terms in the sum); c) the weights of the arcs connected to node $i$ are all maximum. We have then

$$D_1(i) \geq M(n-1)\log_{10}\frac{n-1}{(n-1)^\alpha} = (1-\alpha)M(n-1)\log_{10}(n-1). \tag{4}$$

It is to be noted that conditions a) and b) are met in a fully meshed network for a node whose arcs all exhibit the maximum weight. Though the lower bound implied by Eq (4) depends on $\alpha$, the lower bound is valid when $\alpha = 1$ as well (which leads to $D_1(i) \geq 0$): for that case, since all contributions are non-negative, the lowest we can get is zero, which is achieved either in a full mesh topology or for a terminal node in a star topology (in that case we would have a sum made of a single zero term).

**Distinctiveness centrality**

It is defined as

$$D_2(i) = \sum_{\substack{j=1 \\ j \neq i}}^{n} \log_{10}\frac{n-1}{g_j^\alpha}I_{(w_{ij}>0)}. \tag{5}$$

This metric can be seen as the degree centrality [9] adjusted through the same logarithmic term used in $D_1$. Alternatively, it can be seen as a variant of $D_1$ where arc weights are not considered, but just the number of connections a node has. Mathematically $D_2$ is equal to $D_1$ with $w_{ij}$ = 1. We can then derive the lower and upper bounds for $D_2$ from those obtained for $D_1$ just by setting $M$ = 1 in Eqs (4) and (3). We get

$$D_2(i) \leq (n-1)\log_{10}(n-1), \tag{6}$$

and

$$D_2(i) \geq (1-\alpha)(n-1)\log_{10}(n-1). \tag{7}$$

### Global weight distinctiveness centrality

It is defined as

$$D_3(i) = \sum_{\substack{j=1 \\ j \neq i}}^{n} w_{ij} \log_{10} \frac{\sum_{\substack{k,l=1 \\ k \neq l}}^{n} \frac{w_{kl}}{2}}{\left(\sum_{\substack{k=1 \\ k \neq j}}^{n} w_{jk}^{\alpha}\right) - w_{ij}^{\alpha} + 1}. \qquad (8)$$

Here again, the index is made of a sum of terms, where just the nodes adjacent to the node of interest $i$ are included. Each adjacent node is accounted for through the weight of the arc connecting it to the node of interest. However, that weight is itself weighted by a logarithmic term that introduces a penalization for those nodes that are highly connected and with large arc weights. When $\alpha = 1$, the denominator in the logarithm argument is the sum of the arc weights for the arcs connected to the nodes adjacent to the node of interest, excluding the arc connecting it to the node of interest. The numerator of the logarithm argument is just a normalization factor (the sum of all arc weights in the graph), introduced to consider the proportion of the total weights that is accountable to the connections of node $j$. The major difference with respect to $D_1$ is that the arc weight, rather than the degree, is considered in the penalization factor.

As for the other metrics, we look for the lower and upper bound of the metric.

As to the lower bound, the argument of the logarithm in Eq (8) can be lowered through the following sequence of inequalities:

$$\log_{10} \frac{\sum_{\substack{k,l=1 \\ k \neq l}}^{n} \frac{w_{kl}}{2}}{\left(\sum_{\substack{k=1 \\ k \neq j}}^{n} w_{jk}^{\alpha}\right) - w_{ij}^{\alpha} + 1} \geq \log_{10} \frac{\sum_{\substack{k=1 \\ k \neq i,j}}^{n} w_{jk} + w_{ij}}{\sum_{\substack{k=1 \\ k \neq i,j}}^{n} w_{jk}^{\alpha} + 1}$$

$$\geq \log_{10} \frac{\sum_{\substack{k=1 \\ k \neq i,j}}^{n} w_{jk} + m}{\sum_{\substack{k=1 \\ k \neq i,j}}^{n} w_{jk}^{\alpha} + 1} \geq \log_{10} \frac{(n-2)M + m}{(n-2)M^{\alpha} + 1} \qquad (9)$$

However, the minimum argument that we get from the last inequality may still be larger than 1 (hence the logarithm would be positive). If that's the case, i.e. if the following condition holds:

$$\frac{(n-2)M + m}{(n-2)M^{\alpha} + 1} > 1 \rightarrow (n-2)(M^{\alpha} - M) < m - 1, \qquad (10)$$

a lower bound for $D_3$ is obtained for a sum made of just a single term and the minimum factor:

$$D_3(i) > m \log_{10} \frac{(n-2)M + m}{(n-2)M^{\alpha} + 1}. \qquad (11)$$

On the other hand, if the condition (10) does not hold, the logarithm turns negative, and a lower bound can be obtained by considering the maximum possible number of terms in the sum, with the maximum factor:

$$D_3(i) > (n-1)M \log_{10} \frac{(n-2)M + m}{(n-2)M^\alpha + 1}. \tag{12}$$

It is to be noted that the upper bound in the case of negative terms in the sum has been obtained by acting separately on the logarithm and on the sum itself, though the two actions rely on conditions that may not take place at the same time: the resulting lower bound may then be quite loose.

As to the upper bound, again, it seems not to be possible to get an upper bound as tight as we do for the other metrics. As in the other cases, we may draft a list of the features we wish to maximize $D_3$: a) maximizing the number of terms in the sum (8); b) maximizing the weight $w_{ij}$ of the arc connecting the node of interest to its adjacent nodes; c) maximizing the numerator of the argument of the log; d) minimizing the denominator of the argument of the log. Unfortunately, the terms in the sum are not independent of each other, and increasing the number of terms decreases the value of the individual terms: the aims highlighted above typically conflict with each other.

A loose upper bound can be obtained if we satisfy all the conditions reported above, regardless of their interactions. In particular, we consider a sum made of the maximum possible number of terms, with the maximum factor $w_{ij} = M$. The maximum of the logarithm is considered when the neighbouring nodes have no other connections (to lower the denominator to just the 1 term), and the numerator is computed as for a full mesh with maximum weights (which is the maximum that the sum of all arc weights can get):

$$D_3(i) < (n-1)M \log_{10} \frac{n(n-1)M}{2}. \tag{13}$$

### Weighted proportional distinctiveness centrality

This metrics is defined as

$$D_4(i) = \sum_{\substack{j=1 \\ j \neq i}}^{n} w_{ij} \frac{w_{ij}^\alpha}{\sum_{\substack{k=1 \\ k \neq j}}^{n} w_{jk}^\alpha}. \tag{14}$$

As in the case of $D_3$, this metric is a weighted sum of arc weights, but it differs for the choice of the penalization factor. This factor is now the ratio of the weight of the arc connecting node $i$ to the neighbouring node (raised to the power of $\alpha$) to the sum of weights (raised to the power of $\alpha$) of all the arcs connected to the neighbouring node. The factor penalizes nodes connected to highly connected nodes so that we expect the metric to be large for nodes that are highly connected to nodes that are poorly connected. It is to be noted that this metric is always positive (for non-isolated nodes).

Considering the nodes connected to $i$ (those for which $w_{ij} \geq 1$), we can rewrite the expression of $D_4$ as

$$D_4(i) = \sum_{\substack{j=1 \\ j \neq i}}^{n} \frac{w_{ij}^{\alpha+1}}{w_{ji}^\alpha + \sum_{\substack{k=1 \\ k \neq j,i}}^{n} w_{jk}^\alpha} = \sum_{\substack{j=1 \\ j \neq i}}^{n} \frac{w_{ij}}{1 + \sum_{\substack{k=1 \\ k \neq j,i}}^{n} \left(\frac{w_{jk}}{w_{ij}}\right)^\alpha}, \tag{15}$$

since $w_{ij} = w_{ji}$ for an undirected network.

The maximum of this metric is achieved when the following conditions hold: a) node $i$ is maximally connected ($g_i = n - 1$); b) the arcs of node $i$ have maximum weight $M$; c) the neighbouring nodes are not connected to any other node ($w_{jk} = 0$ for $k \neq i$). These conditions are achieved for the hub node in a star network when all the arcs have weight $M$. We have then

$$D_4(i) \leq (n-1)M. \quad (16)$$

The minimum of the metric is achieved similarly with the dual conditions: a) the node $i$ is connected to a single node; b) the arc from node $i$ has minimum weight $m$; c) the neighbour to node $i$ is connected to all other nodes with maximum weight ($w_{jk} = M$ for $k \neq i$), so that

$$D_4(i) \geq \frac{m}{1 + (n-2)(\frac{M}{m})^\alpha}. \quad (17)$$

## Proportional distinctiveness centrality

This final metric is defined as

$$D_5(i) = \sum_{\substack{j=1 \\ j \neq i}}^{n} \frac{1}{g_j^\alpha} I_{(w_{ij} > 0)}. \quad (18)$$

This metric just considers the reciprocals of the degrees of the nodes adjacent to $i$, raised to the power of $\alpha$. Again, the rationale is that neighbouring poorly connected nodes count more so that the most influential nodes are those connecting poorly connected nodes.

Here we have again a metric made of positive contributions, as for $D_4$, but differently from what happens for the first three metrics. This metric is maximized if we consider a node that is maximally connected, whose neighbouring nodes instead have minimal connectivity ($g_j = 1$). This is the case of the hub in a star network. The upper bound is then:

$$D_5(i) \leq n - 1. \quad (19)$$

As to the lower bound, the same arguments lead us to consider a node that has minimal connectivity ($g_i = 1$), with its only neighbour having maximum degree ($g_j = n - 1$). This is what we have for any terminal node in a star network. The lower bound is therefore obtained by considering a single-term summation:

$$D_5(i) \geq \frac{1}{(n-1)^\alpha}. \quad (20)$$

## A toy application example

In this section, for the purpose of illustrating the computation and the specific features of each metric, we employ a 6-node toy network, shown in Fig 1.

Table 1 shows the different values of the distinctiveness metrics for $\alpha = 1, 2, 5$. We have highlighted in red the highest value of each column and in blue the lowest one. We notice that with $\alpha > 1$ connections to high-degree nodes have a stronger negative impact on centrality, such that their contribution becomes negative for $D_1$, $D_2$ and $D_3$. This does not happen in the case of $D_4$ and $D_5$, for which each additional arc makes a positive contribution, however small.

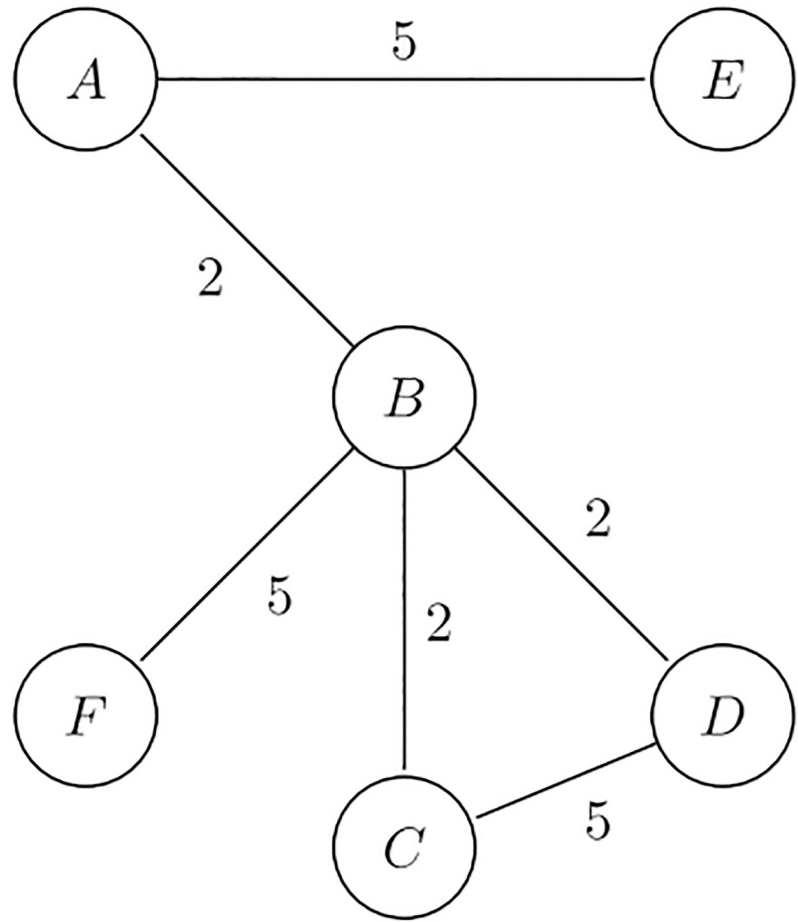

**Fig 1. Toy network.**

https://doi.org/10.1371/journal.pone.0233276.g001

In the network of Fig 1, for $\alpha = 1$, nodes *C* and *D* have a higher distinctiveness centrality than node *E*. However, the centrality ranking changes significantly when we increase $\alpha$. For $D_1$, $D_2$, and $D_3$, nodes *C* and *D* become less central than *E* already at $\alpha = 2$, due to their connection with *B* which is the network hub. It is also important to notice that with $\alpha = 1$ the nodes with the maximum and minimum centrality are the same for all metrics (B and F respectively). However, when $\alpha \geq 1$, these rankings change and may disagree with each other.

Table 2 additionally shows the values of some of the most popular centrality metrics—i.e. non-normalized betweenness, closeness [2] and eigenvector centrality [11]. The values of degree, Burt's constraint and effective size metrics [13, 14] are also reported in the table. All metrics were calculated through the Python Networkx package [21] both for the unweighted network (Table 2a) and the weighted one (Table 2b). In our network, arc weights represent the strength of relationships; inverse weights were used for the computation of network paths where needed.

From a quick comparison of the values reported in the table, we see that the information captured by each metric is different. This is also true for the effective size measure, whose

Table 1. Distinctiveness centrality metrics in the toy network.

| Node | $D_1$ | $D_2$ | $D_3$ | $D_4$ | $D_5$ |
|------|-------|-------|-------|-------|-------|
| A | 3.689 | 0.796 | 7.256 | 5.364 | 1.250 |
| B | 5.882 | 1.893 | 9.876 | 6.714 | 2.500 |
| C | 2.184 | 0.495 | 4.870 | 3.935 | 0.750 |
| D | 2.184 | 0.495 | 4.870 | 3.935 | 0.750 |
| E | 1.990 | 0.398 | 4.225 | 3.571 | 0.500 |
| F | 0.485 | 0.097 | 2.386 | 2.273 | 0.250 |

(a) $\alpha = 1$

| Node | $D_1$ | $D_2$ | $D_3$ | $D_4$ | $D_5$ |
|------|-------|-------|-------|-------|-------|
| A | 2.485 | 0.194 | 6.193 | 5.216 | 1.062 |
| B | 4.076 | 0.990 | 6.055 | 5.828 | 1.750 |
| C | -0.526 | -0.408 | 2.698 | 4.527 | 0.312 |
| D | -0.526 | -0.408 | 2.698 | 4.527 | 0.312 |
| E | 0.485 | 0.097 | 3.116 | 4.310 | 0.250 |
| F | −2.526 | −0.505 | 1.041 | 3.378 | 0.062 |

(b) $\alpha = 2$

| Node | $D_1$ | $D_2$ | $D_3$ | $D_4$ | $D_5$ |
|------|-------|-------|-------|-------|-------|
| A | −1.128 | -1.612 | 2.248 | 5.020 | 1.001 |
| B | -1.342 | -1.720 | −6.426 | 5.061 | 1.094 |
| C | -8.654 | −3.118 | -5.345 | 4.969 | 0.032 |
| D | -8.654 | −3.118 | -5.345 | 4.969 | 0.032 |
| E | -4.031 | −0.806 | -0.981 | 4.949 | 0.031 |
| F | −11.557 | -2.311 | -3.323 | 4.851 | 0.001 |

(c) $\alpha = 5$

https://doi.org/10.1371/journal.pone.0233276.t001

Table 2. Popular centrality metrics in the toy network.

| Node | DG | BTW | CLO | EIG | CON | ES |
|------|----|----|-----|-----|-----|-----|
| A | 2 | 4 | 0.625 | 0.321 | 0.500 | 2.000 |
| B | 4 | 8 | 0.833 | 0.628 | 0.406 | 3.500 |
| C | 2 | 0 | 0.555 | 0.455 | 0.953 | 1.000 |
| D | 2 | 0 | 0.555 | 0.455 | 0.953 | 1.000 |
| E | 1 | 0 | 0.417 | 0.135 | 1.000 | 1.000 |
| F | 1 | 0 | 0.500 | 0.264 | 1.000 | 1.000 |

(a) Unweighted network

| Node | WDG | WBTW | WCLOS | WEIG | WCON | WES |
|------|-----|------|-------|------|------|-----|
| A | 7 | 4 | 0.625 | 0.275 | 0.592 | 2.000 |
| B | 11 | 8 | 0.833 | 0.572 | 0.434 | 3.636 |
| C | 7 | 0 | 0.556 | 0.458 | 0.827 | 1.600 |
| D | 7 | 0 | 0.556 | 0.458 | 0.827 | 1.600 |
| E | 5 | 0 | 0.417 | 0.183 | 1.000 | 1.000 |
| F | 5 | 0 | 0.500 | 0.381 | 1.000 | 1.000 |

(b) Weighted network

DG = degree; WDG = weighted degree; BTW = betweenness; WBTW = weighted betweenness; CLO = closeness; WCLOS = weighted closeness; EIG = eigenvector centrality; WEIG = weighted eigenvector centrality; CON = constraint; WCON = weighted constraint; ES = effective size; WES = weighted effective size.

https://doi.org/10.1371/journal.pone.0233276.t002

conceptualization is based on the concept of redundancy: an ego has redundancy if its contacts are connected to each other as well. Distinctiveness centrality rankings differ from those obtained through degree, closeness, betweenness, eigenvector centrality, effective size and constraint.

## A comparison with established metrics

In order to extend the comparison of distinctiveness centrality with other popular and frequently-used network metrics [2, 9, 11, 14], we generated 1000 random scale-free networks, according to the Barabási–Albert preferential attachment model (with 50 nodes and 2 arcs that are preferentially attached to existing nodes with high degree, when the network grows). We used the Networkx Python package [21]. Weights of existing arcs were assigned through a uniform selection of random integers in the range [1, 20]. As we did in the previous section, we treated arc weights as the strength of relationships.

For each network, we computed the Spearman's rank correlation coefficients for all pairs of metrics and several values of $\alpha$, to see how similar their centrality rankings were. Average correlations are shown in the tables provided as Supporting Information (S2 File), for $\alpha$ = 1, 2 and 5.

We see no perfect overlaps ($\rho$ = 1 or $\rho$ = -1), which means that no two metrics are perfectly interchangeable (i.e. redundant). As expected, rankings produced by our metrics are similar to each other, since they are consistent with the same goal (attributing greater relevance to nodes bridging the network periphery). When $\alpha$ = 1, distinctiveness centrality metrics are most correlated with degree and weighted degree. Increasing the value of $\alpha$ leads to a larger penalization of connections towards high-degree nodes so that correlations with the other indicators drop and sometimes also become negative. For example, if $\alpha$ increases from 1 to 2, the average correlations of $D_1$, $D_2$ and $D_3$ with closeness and eigenvector centrality are nearly halved.

Figs 2 and 3, are more informative and show the average correlations of DC metrics with the other metrics, for more values of $\alpha$. In all plots, $D_1$, $D_2$ and $D_3$ have the correlations that decrease more rapidly, quickly reaching negative scores (for the measure of constraint the effect is of course inverted). On the other hand, rankings obtained through $D_4$ and $D_5$ are the most stable, i.e. they do not change much when $\alpha$ is increased. There are no cases of perfect ranking overlap if we take $\alpha \geq 1$ as in the definition of distinctiveness centrality. In general, all correlations seem to stabilize above specific $\alpha$ thresholds.

## Directed networks

Distinctiveness centrality can be further generalized to consider directed networks, where not every arc is reciprocated, and weights may differ in dyadic relationships. Similarly to the case of in- and out-degree [9], we can calculate distinctiveness centrality on directed graphs, considering the number and weight of arcs pointing to each node. Accordingly, we indicate with $g_i^+$ the out-degree of the generic node $i$ and with $g_i^-$ its in-degree. We also notice that in directed networks the arc originating at node $i$ and terminating at node $j$ has a weight $w_{ij}$ that is potentially different from that of its reciprocal $w_{ji}$.

When conceptualizing DC for directed networks, we want to value incoming arcs more if they originate at nodes with low out-degree. Indeed, a connection from a node sending arcs towards all other nodes is considered of little value. We explain this through an example of love-letter writing. Let us consider the case where student A receives a love-letter from student B, who is sending love-letters to all people in the school. The letter sent to A is much less important to A than the case of B sending only one letter (to A). Indeed, B is 'spamming' all the network, sending many outgoing arcs, then each of them gives a low contribution to the receiver's importance. Similarly, we want to value outgoing arcs more when they reach peers

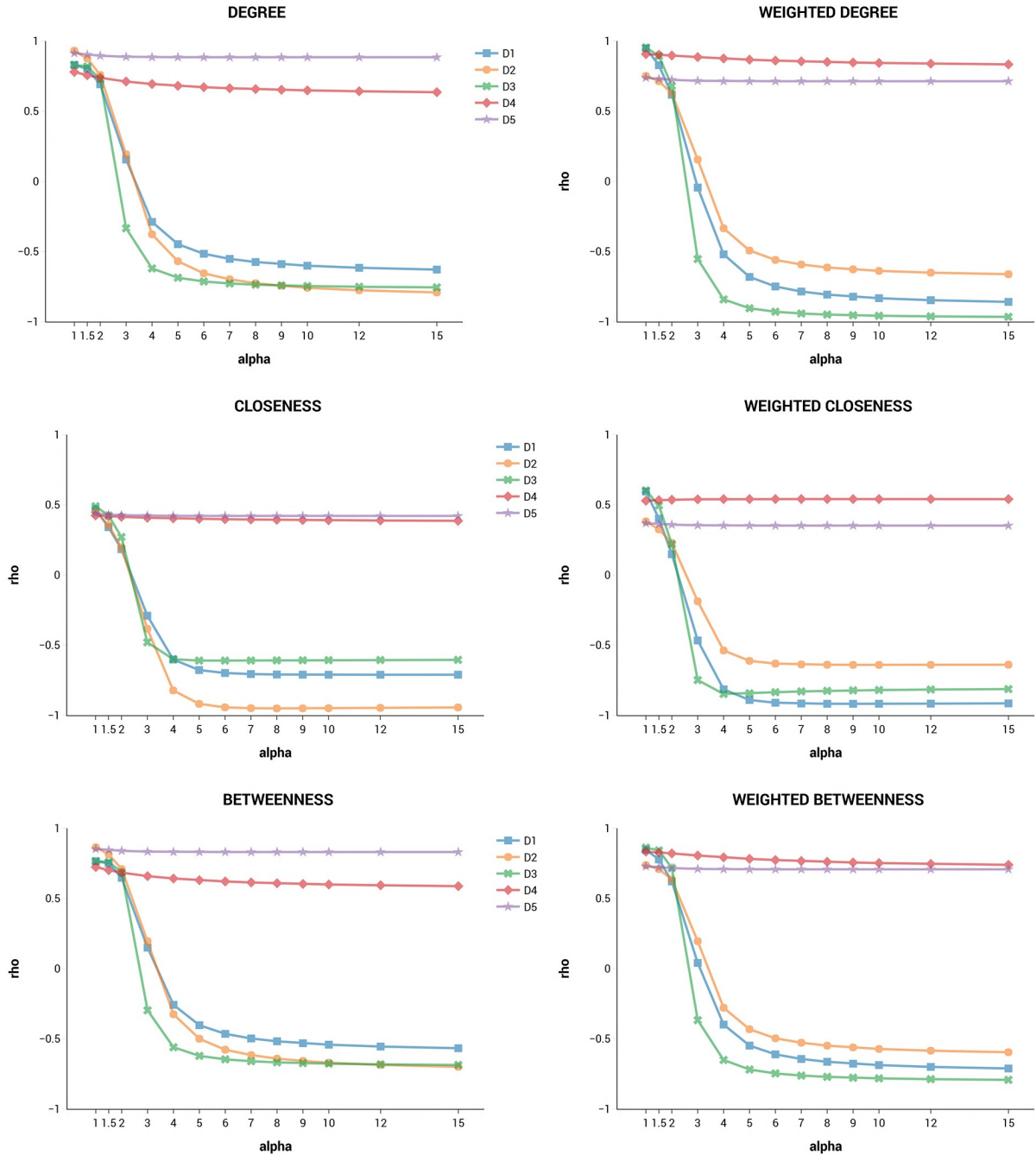

**Fig 2. Spearman's correlation plots of DC with degree, closeness and betweenness.**

https://doi.org/10.1371/journal.pone.0233276.g002

with low in-degree. If the arc sent by a node is the only one, or among the few, to reach another node, that arc will be important. To keep going with our example, if student A is receiving a love letter from student B only, this is much more important than the case of A receiving many love letters. Following this logic, we generalize the equations of distinctiveness centrality to the case of directed networks, thus defining in- and out-distinctiveness:

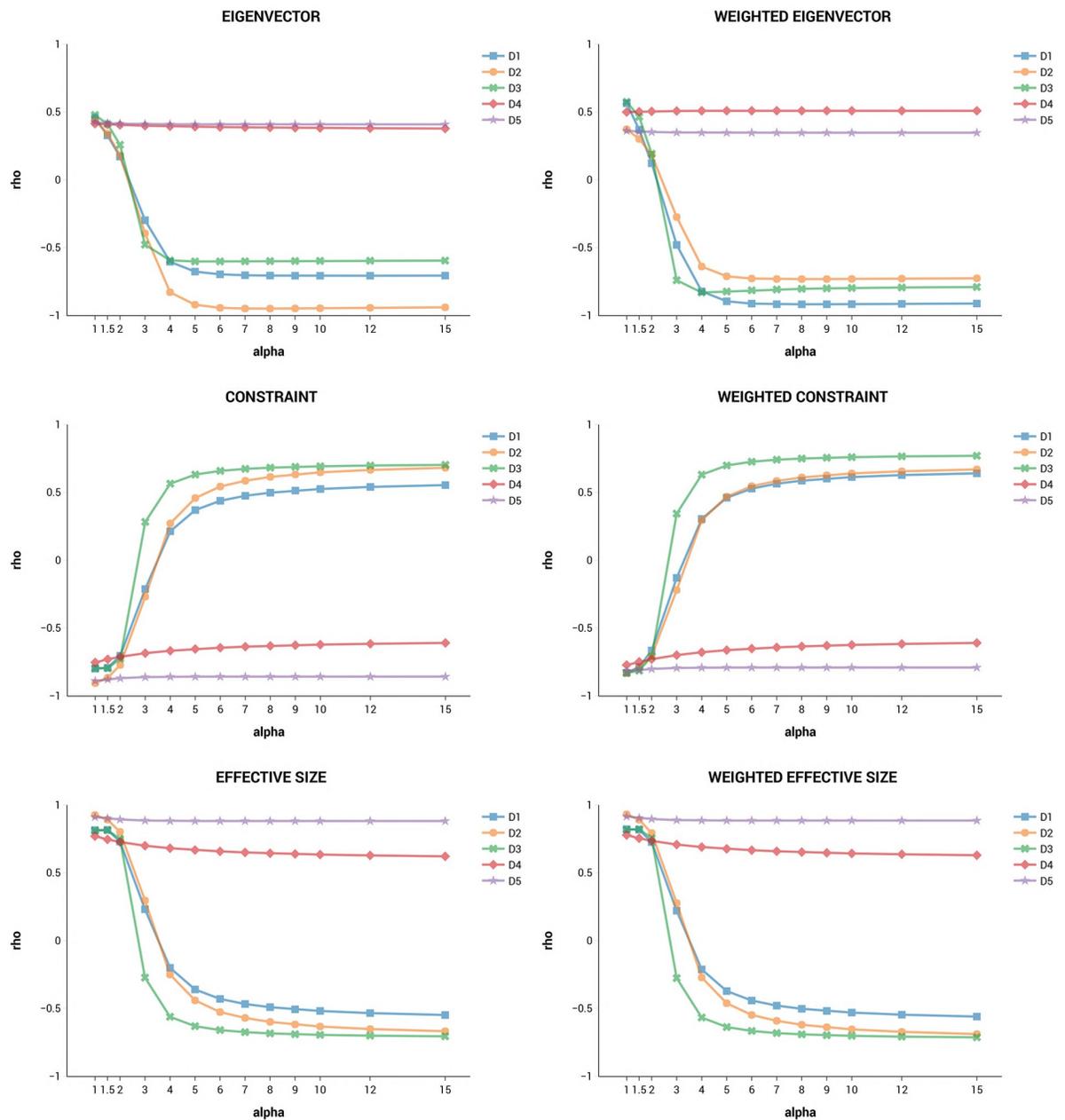

**Fig 3. Spearman's correlation plots of DC with eigenvector centrality, constraint and effective size.**

https://doi.org/10.1371/journal.pone.0233276.g003

- **Weighted Distinctiveness Centrality IN and OUT**

$$D_1^-(i) = \sum_{\substack{j=1 \\ j \neq i}}^{n} w_{ji} \log_{10} \frac{n-1}{g_j^+ \alpha}. \tag{21}$$

$$D_1^+(i) = \sum_{\substack{j=1 \\ j \neq i}}^{n} w_{ij} \log_{10} \frac{n-1}{g_j^- \alpha}. \tag{22}$$

- **Distinctiveness Centrality IN and OUT**

$$D_2^-(i) = \sum_{\substack{j=1 \\ j \neq i}}^{n} \log_{10} \frac{n-1}{g_j^+ \alpha} I_{(w_{ji}>0)}. \tag{23}$$

$$D_2^+(i) = \sum_{\substack{j=1 \\ j \neq i}}^{n} \log_{10} \frac{n-1}{g_j^- \alpha} I_{(w_{ij}>0)}. \tag{24}$$

- **Global Weight Distinctiveness Centrality IN and OUT**

$$D_3^-(i) = \sum_{\substack{j=1 \\ j \neq i}}^{n} w_{ji} \log_{10} \frac{\sum_{\substack{k,l=1 \\ k \neq l}}^{n} w_{kl}}{(\sum_{\substack{k=1 \\ k \neq j}}^{n} w_{jk}^\alpha) - w_{ji}^\alpha + 1}. \tag{25}$$

$$D_3^+(i) = \sum_{\substack{j=1 \\ j \neq i}}^{n} w_{ij} \log_{10} \frac{\sum_{\substack{k,l=1 \\ k \neq l}}^{n} w_{kl}}{(\sum_{\substack{k=1 \\ k \neq j}}^{n} w_{kj}^\alpha) - w_{ij}^\alpha + 1}. \tag{26}$$

- **Weighted Proportional Distinctiveness Centrality IN and OUT**

$$D_4^-(i) = \sum_{\substack{j=1 \\ j \neq i}}^{n} w_{ji} \frac{w_{ji}^\alpha}{\sum_{\substack{k=1 \\ k \neq j}}^{n} w_{jk}^\alpha}. \tag{27}$$

$$D_4^+(i) = \sum_{\substack{j=1 \\ j \neq i}}^{n} w_{ij} \frac{w_{ij}^\alpha}{\sum_{\substack{k=1 \\ k \neq j}}^{n} w_{kj}^\alpha}. \tag{28}$$

- **Proportional Distinctiveness Centrality IN and OUT**

$$D_5^-(i) = \sum_{\substack{j=1 \\ j \neq i}}^{n} \frac{1}{g_j^+ \alpha} I_{(w_{ji}>0)}. \tag{29}$$

$$D_5^+(i) = \sum_{\substack{j=1 \\ j \neq i}}^{n} \frac{1}{g_j^- \alpha} I_{(w_{ij}>0)}. \tag{30}$$

Fig 4 presents a directed toy network to illustrate the use of the metrics for directed networks. Table 3 shows the values of in- and out-distinctiveness centrality for this network when $\alpha$ = 1 and $\alpha$ = 2. We highlighted the highest value of each column in red and the lowest one in blue. Node *B* is certainly important due to its outgoing arcs, as it reaches all other nodes in the network, excepting node *E*. However, if we consider weighted out-degree, node *B* has the same

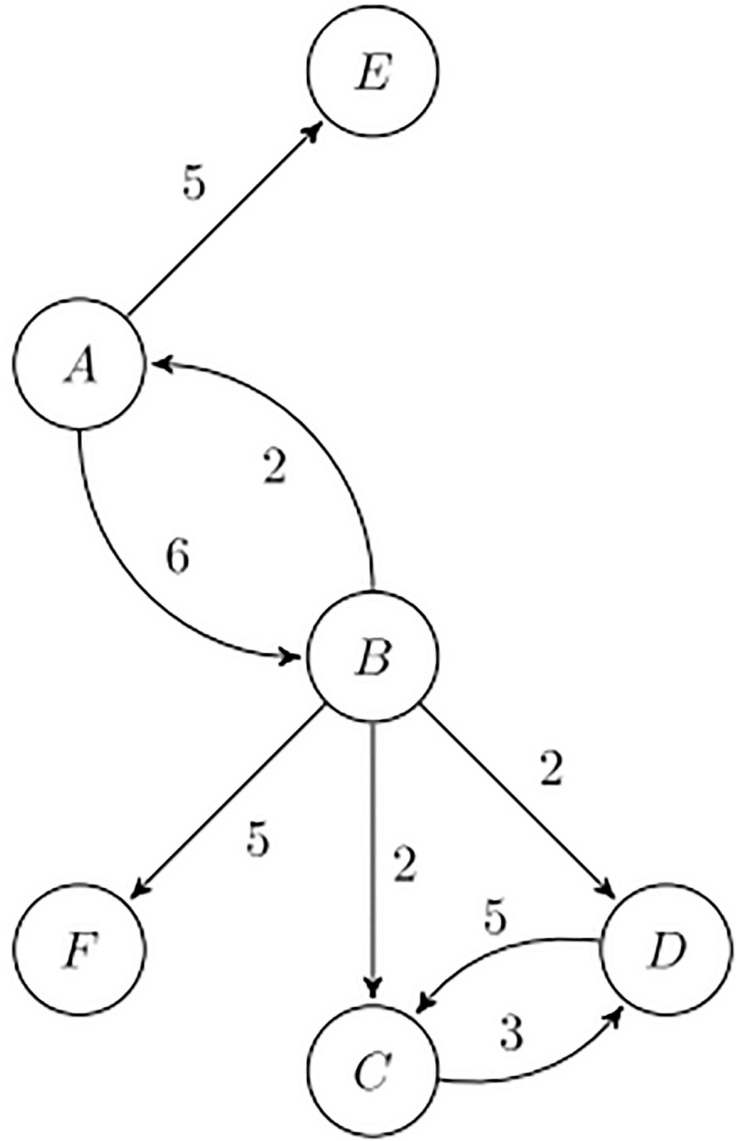

**Fig 4. Directed toy network.**

https://doi.org/10.1371/journal.pone.0233276.g004

score as *A*. Both nodes reach others that would otherwise be isolated (*E* and *F*). The fact that *B* is sending arcs towards other nodes with more incoming connections penalizes its out-distinctiveness score, and makes it less important than *A*, for $D_1^+$, $D_3^+$ and $D_4^+$. This effect is amplified for larger values of *α*. On the other hand, *A* only has an incoming arc of weight 2, originating at *B*, which is a node with high out-degree. This makes *A* less important than all other nodes at *α* = 1, according to $D_1^-$, $D_3^-$, $D_4^-$—and as much important as *F* according to $D_2^-$ and $D_5^-$. If we consider $D_1^-$, $D_2^-$ and $D_5^-$ of nodes *E* and *F*, we see that *F* is lower ranked. Both nodes have a single incoming arc of weight equal to 5, but *F* received this arc from *B*, which is sending links towards many other nodes, thus giving a less relevant contribution to the in-distinctiveness of *F*. On the other hand, node *E* is reached by *A*, which only has two outgoing arcs.

**Table 3. Directed toy network distinctiveness centrality.**

| Node | $D_1^-$ | $D_2^-$ | $D_3^-$ | $D_4^-$ | $D_5^-$ | $D_1^+$ | $D_2^+$ | $D_3^+$ | $D_4^+$ | $D_5^+$ |
|---|---|---|---|---|---|---|---|---|---|---|
| A | 0.194 | 0.097 | 0.954 | 0.364 | 0.250 | 7.689 | 1.398 | 16.248 | 11.000 | 2.000 |
| B | 2.388 | 0.398 | 4.194 | 3.273 | 0.500 | 6.485 | 2.194 | 13.488 | 8.371 | 3.000 |
| C | 3.689 | 0.796 | 8.340 | 5.364 | 1.250 | 1.194 | 0.398 | 3.000 | 1.800 | 0.500 |
| D | 2.291 | 0.796 | 5.386 | 3.364 | 1.250 | 1.990 | 0.398 | 5.000 | 3.571 | 0.500 |
| E | 1.990 | 0.398 | 3.160 | 2.273 | 0.50 | 0.000 | 0.000 | 0.000 | 0.000 | 0.000 |
| F | 0.485 | 0.097 | 3.160 | 2.273 | 0.250 | 0.000 | 0.000 | 0.000 | 0.000 | 0.000 |

(a) $\alpha = 1$

| Node | $D_1^-$ | $D_2^-$ | $D_3^-$ | $D_4^-$ | $D_5^-$ | $D_1^+$ | $D_2^+$ | $D_3^+$ | $D_4^+$ | $D_5^+$ |
|---|---|---|---|---|---|---|---|---|---|---|
| A | -1.010 | −0.505 | -0.109 | 0.216 | 0.062 | 7.689 | 1.398 | 16.248 | 11.000 | 2.000 |
| B | 0.581 | 0.097 | 0.373 | 3.541 | 0.250 | 5.280 | 1.592 | 11.418 | 7.891 | 2.500 |
| C | 2.485 | 0.194 | 7.277 | 5.216 | 1.062 | 0.291 | 0.097 | 2.334 | 2.077 | 0.250 |
| D | 1.087 | 0.194 | 4.323 | 3.216 | 1.062 | 0.485 | 0.097 | 3.891 | 4.310 | 0.250 |
| E | 0.485 | 0.097 | −0.455 | 2.049 | 0.250 | 0.000 | 0.000 | 0.000 | 0.000 | 0.000 |
| F | −2.526 | −0.505 | 1.816 | 3.378 | 0.062 | 0.000 | 0.000 | 0.000 | 0.000 | 0.000 |

(b) $\alpha = 2$

https://doi.org/10.1371/journal.pone.0233276.t003

## Possible applications

Our metrics could have several applications and offer perspectives for future research, including, e.g., the identification of prominent nodes in criminal organizations. These are sometimes organized as groups of semi-independent, or entirely separated small cells, with the absence of large network hubs [22]. In such a scenario, distinctiveness centrality could effectively serve the identification of nodes that keep the network periphery together. Our metrics could also complement information obtained through other approaches. For example, they could be used to test new network fragmentation strategies meant to contain epidemics [23].

In the field of Semantic Network Analysis, Fronzetti Colladon [20] recently presented the Semantic Brand Score (SBS), a measure of brand importance which is computed from the analysis of potentially-big textual data. While it does not fall within the scope of this paper to discuss the construct of brand importance, we maintain that our distinctiveness centrality metric (namely $D_2$) could be considered as an alternative to degree centrality for the measurement of Diversity (one of the components of the SBS). Indeed, we compute the SBS through a network of co-occurring words, where nodes are words appearing in the analysed texts, and links between them are determined by the frequency of their co-occurrences. For example, if the sentence "it is a beautiful day" appears 7 times, the word nodes "beautiful" and "day" will be connected by an arc of weight 7. In this context, the SBS dimension of Diversity counts how many different textual associations exist for each node, and in particular for those nodes that are considered "brands" in the analysis. Diversity was operationalized through degree centrality [2], without penalizing the connections of the brand node to high-degree nodes. In our view, it could be useful to distinguish brands with common textual associations (shared with many other nodes) from brands that have more exclusive relationships with specific words. To this purpose, distinctiveness centrality ($D_2$) could be considered as a reasonable candidate. The idea of adjusting the SBS Diversity metric is also aligned with the logic behind the term frequency—inverse document frequency (TF-IDF) normalization process that is very often used in text analysis [24, 25]. According to Robertson [26] words within a document can be divided in those with eliteness and those without. TF-IDF helps understanding how important a word is to a document, which is part of a corpus. Specifically, we can consider the DTM (Document

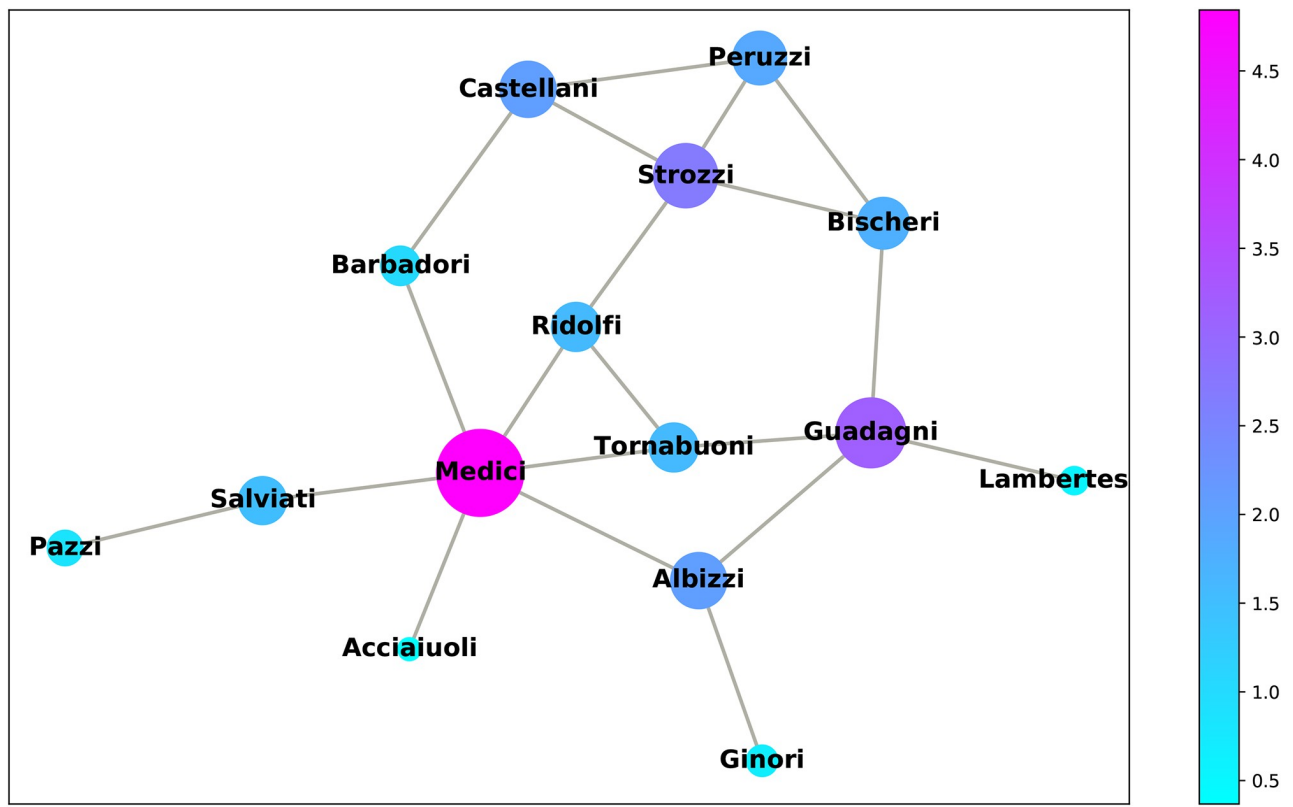

**Fig 5. Florentine families in the 15th century, with colour and size according to $D_1$ ($\alpha = 1$).**

https://doi.org/10.1371/journal.pone.0233276.g005

Term Matrix) of the corpus, where documents make up rows, and words make up the columns in the corpus. This matrix is populated by values that reflect the frequency of appearance of each word in each document. However, frequency is not sufficient to understand the word-importance to a document—as well as Prevalence is not sufficient to define the SBS. There might be words, such as "and", which add little meaning to the discourse and appear with high frequency in all documents. In order to identify distinctive words, we transform frequency values into TF-IDF values, which increase proportionally to the number of times a word appears in a document and are offset by the number of documents in the corpus that contain that word. This is what $D_2$ and our other distinctiveness centrality metrics do: they attribute more importance to the links that more strongly connect a node with low-degree peers; in the case of a word network, strong links to distinctive words are privileged.

In the following, we provide two more examples based on the analysis of two popular real-world networks: the first (Fig 5) is the unweighted network of marital relationships between Florentine families in the 15th century (available on Networkx); the second (Fig 6) is the weighted network of the Zachary's karate club [27] (downloaded from the accompanying material of the book of Latora and colleagues [28], https://www.complex-networks.net/datasets.html).

We are interested in comparing the rankings obtained through DC and the other metrics considered so far. These are shown in Table 4 for the first network. Here, we use $D_1$, $D_3$ and $D_5$ as metrics of distinctiveness, because in unweighted networks—were $w_{ij} = 1$ for all existing arcs—there is no difference between $D_1$ and $D_2$ and between $D_4$ and $D_5$. In the table we used, as an example, two different values of $\alpha$ ($\alpha = 1, 2$), omitting $D3$ for $\alpha = 2$ as this metric (mainly

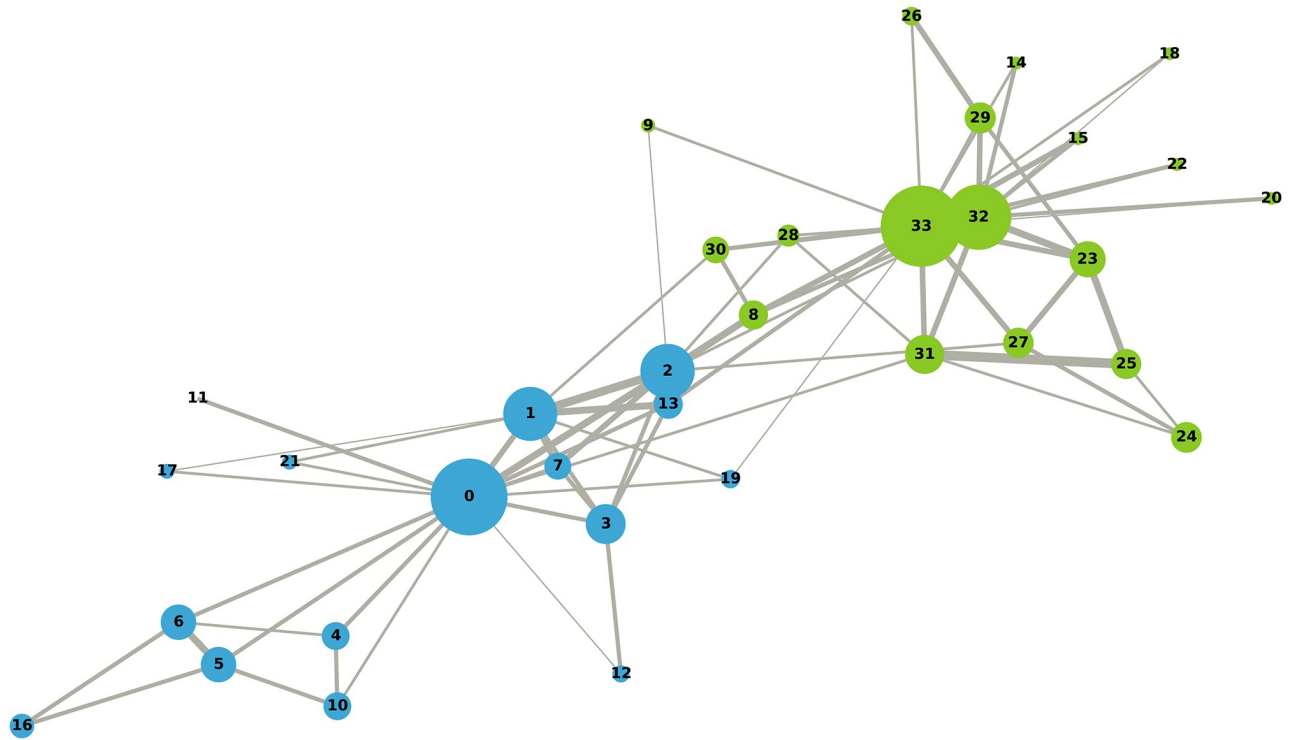

**Fig 6. Zachary's karate club network.**

https://doi.org/10.1371/journal.pone.0233276.g006

**Table 4. Node ranking: Florentine families.**

| Family | DG | BETW | CLOS | EIG | CON | ES | $D_1$ | $D_3$ | $D_5$ | $D_1$ | $D_5$ |
| --- | --- | --- | --- | --- | --- | --- | --- | --- | --- | --- | --- |
| | | | | | | | $\alpha = 1$ | | | $\alpha = 2$ | |
| Medici | 1 | 1 | 1 | 1 | 15 | 1 | 1 | 1 | 1 | 1 | 1 |
| Guadagni | 2 | 2 | 5 | 5 | 14 | 2 | 2 | 2 | 2 | 2 | 2 |
| Strozzi | 2 | 7 | 6 | 2 | 12 | 3 | 3 | 3 | 4 | 3 | 5 |
| Albizzi | 4 | 3 | 3 | 9 | 13 | 3 | 4 | 4 | 3 | 6 | 3 |
| Bischeri | 4 | 6 | 8 | 6 | 8 | 5 | 7 | 7 | 8 | 10 | 9 |
| Castellani | 4 | 10 | 9 | 8 | 8 | 5 | 4 | 5 | 6 | 5 | 6 |
| Peruzzi | 4 | 11 | 11 | 7 | 5 | 11 | 6 | 6 | 7 | 8 | 7 |
| Ridolfi | 4 | 5 | 2 | 3 | 10 | 5 | 8 | 8 | 9 | 13 | 10 |
| Tornabuoni | 4 | 9 | 3 | 4 | 10 | 5 | 8 | 8 | 9 | 13 | 10 |
| Barbadori | 10 | 8 | 6 | 10 | 6 | 9 | 11 | 11 | 11 | 12 | 12 |
| Salviati | 10 | 4 | 9 | 11 | 6 | 9 | 10 | 10 | 5 | 4 | 4 |
| Acciaiuoli | 12 | 12 | 11 | 12 | 1 | 12 | 15 | 15 | 15 | 15 | 15 |
| Ginori | 12 | 12 | 13 | 14 | 1 | 12 | 13 | 13 | 13 | 9 | 13 |
| Lamberteschi | 12 | 12 | 14 | 13 | 1 | 12 | 14 | 14 | 14 | 11 | 14 |
| Pazzi | 12 | 12 | 15 | 15 | 1 | 12 | 12 | 12 | 11 | 7 | 8 |

DG = degree; BTW = betweenness; CLO = closeness; EIG = eigenvector centrality; CON = constraint; ES = effective size.

https://doi.org/10.1371/journal.pone.0233276.t004

Table 5. Spearman's correlation coefficients for the Zachary's karate club network.

| Measure | DG | WDG | WBETW | WCLOS | WEIG | WCON | WES |
|---|---|---|---|---|---|---|---|
| $\alpha = 1$ | | | | | | | |
| $D_1$ | 0.928 | 0.974 | 0.824 | 0.711 | 0.767 | −0.834 | 0.906 |
| $D_2$ | 0.930 | 0.925 | 0.849 | 0.742 | 0.695 | −0.835 | 0.933 |
| $D_3$ | 0.931 | 0.984 | 0.829 | 0.715 | 0.787 | −0.830 | 0.900 |
| $D_4$ | 0.887 | 0.966 | 0.786 | 0.643 | 0.739 | −0.780 | 0.865 |
| $D_5$ | 0.896 | 0.901 | 0.824 | 0.696 | 0.637 | −0.794 | 0.906 |
| $\alpha = 2$ | | | | | | | |
| $D_1$ | 0.278 | 0.332 | 0.295 | 0.042 | -0.063 | −0.152 | 0.376 |
| $D_2$ | 0.426 | 0.518 | 0.413 | 0.165 | 0.154 | −0.319 | 0.517 |
| $D_3$ | 0.901 | 1.959 | 0.812 | 0.677 | 0.721 | −0.792 | 0.886 |
| $D_4$ | 0.858 | 0.951 | 0.769 | 0.606 | 0.725 | −0.746 | 0.839 |
| $D_5$ | 0.854 | 0.873 | 0.787 | 0.634 | 0.581 | −0.740 | 0.872 |

DG = degree; WDG = weighted degree; WBTW = weighted betweenness; WCLOS = weighted closeness; WEIG = weighted eigenvector centrality; WCON = weighted constraint; WES = weighted effective size.

https://doi.org/10.1371/journal.pone.0233276.t005

conceived for weighted networks) does not change when $\alpha$ increases, if all arc weights are equal to 1. We can see that the rankings calculated through DC never overlap with the others in the table, proving that our metrics capture different information. The Medici family is ranked first for all metrics (including constraint, for which we have to consider the inverse ranking), and the Guadagni family ranks second (except for closeness and eigenvector centrality). The metric $D_1$ (for both values of $\alpha$) and $D_3$ (for $\alpha = 1$) both rank the Strozzi family third (apart from them, this only happens for the measure of effective size, which however ranks the Strozzi and Albizzi families equally). If we take the first three families together (as ranked by $D_1$ and $D_3$ for $\alpha = 1$), we can reach all other nodes in the network with a direct connection, only excluding the more peripheral Pazzi and Ginori. At $\alpha = 2$ we see that the $D_1$ ranking of the Albizzi family is lower, due to its links with the Guadagni's and Medici's families that are highly connected.

In the second example, we computed the DC of the 34 members of the Zachary's karate club [27]. Fig 6 shows their friendship network, based on a two-year observation of their relationships. Arc weights represent the number of different contexts in which two individuals interacted. Due to a conflict between the club administrator (node 0) and the instructor (node 33), the club split into two (with the two partitions represented in the figure with different colours). Table 5 shows the Spearman's correlation coefficients of distinctiveness centrality (computed for $\alpha = 1, 2$) with the weighted version of the other metrics. We highlighted in red the highest value of each row and in blue the lowest one. Again, we find no information overlap, and correlations decrease when $\alpha$ increases. Some correlations drop faster than others while increasing $\alpha$, and this also depends on the network structure. In this case, $D_1$ and $D_2$ are the measures that exhibit the fastest decrease.

Nodes are coloured by partition and their size varies according to $D_2$ ($\alpha = 1$), with bigger nodes indicating higher values of the metric. Thicker arcs indicate stronger relationships.

## Discussion and conclusions

The conceptualization of distinctiveness centrality contributes to network theory and introduces a new perspective in network studies. The set of distinctiveness centrality metrics we have presented in this paper could be used in multiple contexts—in all cases where it is important to value the role of nodes connecting low-degree peers. Those nodes have more distinctive connections and are often a bridge to reach the network periphery. We have additionally evaluated the upper and lower bound of each metric.

As shown in the Definition of metrics section, the node influence measured by distinctiveness centrality is different from that measured by degree, weighted degree, closeness, eigenvector, and betweenness centrality, Burt's [13, 14] constraint and effective size. The information captured by our metrics is different. We found that Spearman's correlation coefficients of distinctiveness centrality with popular centrality and ego-network metrics decrease as $\alpha$ increases—even reaching high negative correlations in some cases, for high values of $\alpha$. This property was tested on random scale-free networks [29, 30], where we found no perfect overlap of rankings with degree, closeness, betweenness, eigenvector centrality, effective size and Burt's constraint [13] (both in their weighted and unweighted versions). Spearman's correlation coefficients of these metrics with DC were never equal to 1 or -1. When $\alpha$ was bigger than 1, such correlations could become negative. This happened faster for $D_1$, $D_2$ and $D_3$. On the other hand, correlations with $D_4$ and $D_5$ remained more stable and always positive (excepting Burt's constraint, for which the correlations are inverted).

This paper has the goal of defining distinctiveness centrality and presenting its main properties. We have also discussed possible applications and provided some preliminary examples based on the analysis of well-known real-world networks. However, dedicated research is needed to dig deeper into the possible applications of DC. Indeed, there might be cases in social network analysis where the most important nodes are those that keep together the network periphery, regardless of strong connections with hubs. Analysts could be interested in assessing how exclusive are the connections of some nodes, like in the case of sending and receiving love messages, mentioned earlier in the paper. Even if not within the scope of this paper, we imagine distinctiveness centrality could serve the identification of social actors with many peripheral connections in sparse local communities, that however have no strong relationships with central authorities. Reaching these actors could help strengthen the relationship of local clusters with central authorities—for example, for goals of social inclusion, or to plan interventions to reduce substance abuse [31, 32]. Similarly, individuals with high distinctiveness could be local leaders in covert networks [33], and our metrics could potentially support their identification. These are just some out of many possible hypotheses that could be tested in future studies.

Future research could further explore the properties of our newly defined centrality indicators on network topologies other than scale-free and using different arc weighting approaches. For example, core-periphery structures [34] could be considered, to see how the nodes in the densely connected core are ranked, taking into account that DC penalizes connections with highly-connected peers. The scores and rankings produced by DC metrics could be compared with those obtained through other centrality measures, also considering directed graphs—for example, by comparing with the measures of hub and authority [35].

In order to facilitate the calculation of distinctiveness centrality, we have created a Python package that is freely available at this link https://pypi.org/project/distinctiveness/. We have uploaded its open-source code onto GitHub, with examples and tutorials (https://github.com/iandreafc/distinctiveness). In the future, we plan to provide more free resources, for example, packages written using other programming languages.

## Supporting information

**S1 File. Randomly generated network graphs.** Networks generated using the Python Networkx package [21], according to the procedure presented in the section named "A comparison with established metrics". Files are in the Gexf format.
(ZIP)

**S2 File. Correlations of distinctiveness centrality with well-known centrality and ego-network metrics, calculated on the S1 File.**
(PDF)

## Author Contributions

**Conceptualization:** Andrea Fronzetti Colladon, Maurizio Naldi.

**Data curation:** Andrea Fronzetti Colladon, Maurizio Naldi.

**Formal analysis:** Andrea Fronzetti Colladon, Maurizio Naldi.

**Funding acquisition:** Andrea Fronzetti Colladon, Maurizio Naldi.

**Investigation:** Andrea Fronzetti Colladon, Maurizio Naldi.

**Methodology:** Andrea Fronzetti Colladon, Maurizio Naldi.

**Software:** Andrea Fronzetti Colladon, Maurizio Naldi.

**Validation:** Andrea Fronzetti Colladon, Maurizio Naldi.

**Visualization:** Andrea Fronzetti Colladon, Maurizio Naldi.

**Writing – original draft:** Andrea Fronzetti Colladon, Maurizio Naldi.

**Writing – review & editing:** Andrea Fronzetti Colladon, Maurizio Naldi.